\documentclass[%
 aip,
 amsmath,amssymb,
 reprint,%
]{revtex4-1}

\usepackage{graphicx}
\usepackage{dcolumn}
\usepackage{bm}

\usepackage[utf8]{inputenc}
\usepackage[T1]{fontenc}
\usepackage{mathptmx}
\usepackage{etoolbox}
\usepackage{color}
\usepackage{soul}

\makeatletter
\def\@email#1#2{%
 \endgroup
 \patchcmd{\titleblock@produce}
  {\frontmatter@RRAPformat}
  {\frontmatter@RRAPformat{\produce@RRAP{*#1\href{mailto:#2}{#2}}}\frontmatter@RRAPformat}
  {}{}
}%
\makeatother
\begin{document}

\preprint{AIP/123-QED}
\title[Co-molar volume]{A new thermodynamic function for binary mixtures: the co-molar volume}
\author{Kristian Polanco Olsen}
\affiliation{PoreLab, Department of Physics, Norwegian University of Science and Technology, N-7491 Trondheim, Norway}
\author{Bj{\o}rn Hafskjold}
\email{Bjorn.Hafskjold@ntnu.no}
\affiliation{PoreLab, Department of Chemistry, Norwegian University of Science and Technology, N-7491 Trondheim, Norway}
\author{Anders Lervik}
\email{Anders.Lervik@ntnu.no}
\affiliation{Department of Chemistry, Norwegian University of Science and Technology, N-7491 Trondheim, Norway}
\author{Alex Hansen}
\email{Alex.Hansen@ntnu.no}
\affiliation{PoreLab,Department of Physics, Norwegian University of Science and Technology, N-7491 Trondheim, Norway}
\date{\today}
\begin{abstract}
We have developed a new theory relating partial molar volumes of binary mixtures to the specific (Voronoi) volumes. A simple relation gives new insight into the physical meaning of partial molar volumes in terms of the actual volumes occupied by the molecules. Partial molar quantities are defined through the use of the Euler theorem for homogeneous functions. These properties have been in use for a long time, despite the fact that they do not give an intuitive picture of the properties they are to represent. For instance, the \textit{partial molar volume} of a given component in a mixture is the change in the \textit{total} volume with a change in composition, hence it represents the \textit{derivative} of a volume. The molar volume is a measurable property in the laboratory, and as such a body of thermodynamics, but the derived partial molar volume is not a direct measure of the physical volume occupied by the added component. On the other hand, the physical volume can be computed using \textit{e.g.} molecular dynamics simulations by Voronoi tesselation. We shall call this volume the \textit{specific volume}. To bridge the partial molar volume and the specific volume, we define a single new thermodynamic variable - the \textit{co-molar volume} --- thus bringing the latter into thermodynamics. We demonstrate this bridge through molecular dynamics simulations. The co-molar volume is closely related to the co-moving velocity defined in immiscible two-phase flow in porous media.    
\end{abstract}
\maketitle
\section{Introduction}
\label{sec:intro}

An important part of thermodynamics deals with the mixture of miscible fluids \cite{atkins2014physical}.  
The volume of a mixture will typically not be equal to the sum of the volumes of the pure components.  
It is natural to ask what is the contribution to the total volume from each component.  
Thermodynamics, being a continuum theory, has so far not been able to answer this question.  
It can, however, answer the question of how much the \textit{total} volume of the mixture changes when a component is added or removed; these are the {\it partial molar volumes.\/}  However, they are {\it not\/} the volumes occupied by each component.

With the advent of computer simulations such as molecular dynamics (MD) \cite{frenkel2023understanding}, it {\it is\/} now possible to answer the question just posed: what is the volume occupied by each component in a mixture?  
The {\it specific volumes\/} obtained this way have so far not been expressed as thermodynamic functions and incorporated into the thermodynamic formalism.  

We present here a new thermodynamic function, the  {\it co-molar volume\/} which provides a two-way mapping between the partial molar volumes gotten from thermodynamics and the specific volumes computed in simulations. This has as a consequence that knowledge of the volume and the co-molar volume gives the specific volumes, whereas knowledge of the volume alone gives the partial molar volumes. 

We show through examples that the co-molar volume has a simple functional form characterized by few parameters that for a binary mixture are functions of the pressure and temperature alone.
To determine these parameters, the co-molar volume, and the mapping between the partial molar and specific volumes, we use MD simulations of binary Lennard-Jones/spline (LJs) mixtures \cite{hafskjold2019thermodynamic}, chosen for their simple atomistic structure and wide range of behaviors.

The co-molar volume is closely related to the {\it co-moving velocity\/} where it forms part of a thermodynamics-like description of immiscible two-phase flow in porous media \cite{hansen2018relations,roy2020flow,roy2022co,hansen2023statistical,pedersen2023parameterizations,
alzubaidi2024impact,hansen2024linearity,feder2022physics}.
In that context, it makes it possible to calculate the average flow velocity of each of the two immiscible fluids from a knowledge of the total average flow velocity and the co-moving velocity.   

\section{Theoretical background}
\label{sec:theory}

For simplicity, we shall consider a two-component mixture that is miscible in all compositions.
The Gibbs free energy for this system is
\begin{equation}
\label{eq1}
G(T,p,N_1,N_2)= N_1 \mu_1 + N_2 \mu_2 \;,
\end{equation}
where $T$ and $p$ are the temperature and pressure, respectively, $N_1$ and $N_2$ are the mole numbers, and $\mu_1$ and $\mu_2$ are the chemical potentials of the two components.
The volume of the mixture is given by
\begin{equation}
\label{eq2}
V(T,p,N_1,N_2)=\left(\frac{\partial G}{\partial p}\right)_{T,N_1,N_2}\;,
\end{equation}
where the subscripts $T$ and $N_i$ denote that these quantities are held constant during the derivation.

\subsection{Partial molar volumes}
\label{sec:partial_molar}

The \textit{partial molar volumes} of the two components in a binary mixture are defined as
\begin{eqnarray}
{\hat v}_1=\left(\frac{\partial V}{\partial N_1}\right)_{T,p,N_2}\;,\label{eq4}\\
{\hat v}_2=\left(\frac{\partial V}{\partial N_2}\right)_{T,p,N_1}\;.\label{eq5}
\end{eqnarray}
To simplify the notation, we shall in the following assume that these properties are functions of $T$ and $p$ without being explicitly shown. The molar volume, $v = V/N$, can be expressed as
\begin{equation}
\label{eq6a}
v = x_1 \hat{v}_1 + x_2 \hat{v}_2 = 
x_1 ( \hat{v}_1 - \hat{v}_2) + \hat{v}_2  \;.
\end{equation}
We define the \textit{partial molar volume difference},
\begin{equation}
\label{eq7a}
w(x_1) = \hat{v}_1 - \hat{v}_2 =   \left(\frac{\partial v}{\partial x_1}\right)_{T,p}\;.
\end{equation}
By rearranging equation (\ref{eq6a}) and using equation (\ref{eq7a}), we see that
\begin{eqnarray}
 \hat v_1(x_1)&=& v(x_1)+x_2w(x_1)
\label{eq8a}\;,\\
\hat v_2(x_1)&=& v(x_1) - x_1 w(x_1)\label{eq9a} \;.
\end{eqnarray}
We recognize equations (\ref{eq8a}) and (\ref{eq9a}) as Legendre transforms of the molar volume with respect to $w$. 
Hence, $w$ is the natural variable in a thermodynamic sense for the partial molar volumes, not $x_1$. 

We have determined molar volumes for three LJs mixtures using MD simulations for several compositions at the same $T$ and $p$, fitting Redlich-Kister models \cite{prausnitz1998} to the results, and computed $w(x_1)$ by equation (\ref{eq7a}).
The partial molar volumes follow by equations (\ref{eq8a}) and (\ref{eq9a}).

It is convenient to express the molar volume of a binary mixture as the sum of an ideal volume of mixing and an excess volume:
\begin{equation}
v(x_1,x_2) = v^{id}(x_1,x_2) + v^{ex}(x_1,x_2)
\end{equation}
where $v^{id}(x_1,x_2)$ is a linear combination of the molar volumes of the pure components, $v^{id} = x_1 v_1^* + x_2 v_2^*$.
For many mixtures, the excess molar volume can be successfully expanded in powers of mole fractions, the Redlich-Kister (RK) expansion \cite{prausnitz1998}:
\begin{equation}
v^{ex}(x_1,x_2) = x_1 x_2 \sum_{n=0}^\infty A_n(T,p) (x_1 - x_2)^n \;.
\label{eq5001}
\end{equation}
The coefficients $A_n(T,p)$ are known from experiments.

The partial molar volume difference can likewise be written as an ideal term plus an excess term:
\begin{equation}
w(x_1) = w^{id}(x_1) + w^{ex}(x_1) = v_1^* - v_2^* +  \left(\frac{\partial v^{ex}}{\partial x_1}\right)_{T,p}\;,
\end{equation}

Using Eq. (\ref{eq5001}), the RK expansion for $w^{ex}(x_1)$ gives
\begin{equation}
w^{ex}(x_1) = \sum_{n=0}^\infty A_n [2 n x_1 x_2-(x_1 - x_2)^2](x_1 - x_2)^{(n-1)}
\label{eqwex}
\end{equation}
%
In particular, we shall need $w(0)=w(x_1=0)$ and $w(1)=w(x_1=1)$:
\begin{align}
w(0) =& v_1^* - v_2^* +\sum_{n=0}^\infty (-1)^n A_n \nonumber \\
w(1) =& v_1^* - v_2^* - \sum_{n=0}^\infty A_n \nonumber \\
w(1) - w(0) =& - \sum_{n=0}^\infty \left [1 + (-1)^n \right ] A_n
\label{eq5003}
\end{align}

\subsection{Specific volumes}
\label{sec:direct_molar}

We may define the {\it specific volumes\/} 
\begin{eqnarray}
v_1=\frac{V_1}{N_1}\;,\label{eq7}\\
v_2=\frac{V_2}{N_2}\;,\label{eq8}
\end{eqnarray}
where $V_1$ and $V_2$ are the volumes occupied by each of the two components.
These volumes are not measurable, but they can be computed by simulations.
We have used Voronoi tesselation \cite{okabe2000definitions,medvedev2002approach,kadtsyn2022volumetric} with data from MD simulations to quantify the specific volumes.

The molar volume of the mixture is, expressed with the specific volumes,
\begin{equation}
\label{eq9}
v = x_1 v_1 + x_2 v_2\;.
\end{equation}
The similarity between equations (\ref{eq6a}) and (\ref{eq9}),
\begin{equation}
\label{eq11}
v = x_1{\hat v}_1+x_2{\hat v}_2 = x_1v_1+x_2v_2\;,
\end{equation}
does {\it not\/} imply that the partial molar volumes and specific volumes are equal.

\subsection{The co-molar volume}
\label{sec:co-molar}

The most general relation between $(v_1,v_2)$ and $(\hat v_1,\hat v_2)$ is
\begin{eqnarray}
v_1={\hat v}_1-x_2 v_m\;,\label{eq18}\\
v_2={\hat v}_2+x_1 v_m\;,\label{eq19}
\end{eqnarray}
where $v_m$ is a new function, the {\it co-molar volume.\/}
Combining equations (\ref{eq18}) and (\ref{eq19}) and using the definition of $w$, equation (\ref{eq7a}), gives
\begin{equation}
\label{eq19a}
v_m=({\hat v}_1-{\hat v}_2) - (v_1 - v_2) = w - (v_1 - v_2) \;.
\end{equation}
The $v_m$ has the interesting property that, in addition to depending on $v_1$, $v_2$, and $w$, it can be shown to depend \textit{only} on $v_1$ and $v_2$.
This can be seen as follows:
First, we use the definitions of $\hat{v}_1$ and $\hat{v}_2$ (equations (\ref{eq4}) and (\ref{eq5})) and observe that
\begin{equation}
\label{eq3002a}
x_1 \left(\frac{\partial \hat{v}_1}{\partial x_1} \right )_{T,p} + x_2 \left(\frac{\partial \hat{v}_2}{\partial x_1} \right )_{T,p} = 0 \;.
\end{equation}
Second, we differentiate equations (\ref{eq18}) and (\ref{eq19}) with respect to $x_1$:
\begin{eqnarray}
\frac{\partial v_1}{\partial x_1} = \frac{\partial \hat{v}_1}{\partial x_1} - x_2 \frac{\partial v_m}{\partial x_1} + v_m \label{eq30}
\\
\frac{\partial v_2}{\partial x_1} = \frac{\partial \hat{v}_2}{\partial x_1} + x_1 \frac{\partial v_m}{\partial x_1} + v_m \label{eq31}
\end{eqnarray}
Multiplying equation (\ref{eq30}) with $x_1$, equation (\ref{eq31}) with $x_2$ and adding the two equations while using equation (\ref{eq3002a}) gives
\begin{equation}
\label{eq3002b}
v_m = x_1 \left(\frac{\partial v_1}{\partial x_1} \right )_{T,p} + x_2 \left(\frac{\partial v_2}{\partial x_1} \right )_{T,p}
\,.
\end{equation}

The co-molar volume is a function of $(T,p,x_1)$ as $(v_1,v_2)$ and $(\hat v_1,\hat v_2)$ are.
We have that $\hat v_i = v_i$, $i=1,\ 2$, only if $v_m=0$.

Back to the relation between $v_m$ and $w$: combining equations (\ref{eq9}) and (\ref{eq19a}) gives
\begin{eqnarray}
v_1=v + x_2 ( w - v_m) \;,\label{eq129a} \\
v_2=v - x_1 ( w - v_m) \;.\label{eq129b}
\end{eqnarray}
We see that $w$ is a natural variable also for $v_m$ together with $T$ and $p$, $v_m=v_m(T,p,w)$.
Furthermore, knowing the relation between $v_m$ and $w$, Eqs. (\ref{eq18}) and (\ref{eq19}) can be used to convert both ways between the specific volumes and the partial molar volumes.
We will in the following show that a simple approximation gives a particularly convenient relation between $v_m$ and $w$.

\subsection{Series expansion of the co-molar volume}
\label{sec:expansion}

We expand the co-molar volume as a Taylor series in $w$:
\begin{equation}
\label{eq130f}
v_m = \sum_{k=0}^\infty a_k w^k
\;.
\end{equation}
Inspired by the work on two-phase flow in porous media \cite{hansen2018relations,roy2020flow,roy2022co,hansen2023statistical,pedersen2023parameterizations,
alzubaidi2024impact,hansen2024linearity,feder2022physics}, we now \textit{assume} that $v_m$ is approximated with a linear function of $w$:
\begin{equation}
\label{eq130c}
v_m \approx a + b w\;.
\end{equation}
In Section \ref{sec:fin}, we shall examine this approximation.
Using the linear approximation in Eqs. (\ref{eq129a}) and (\ref{eq129b}) gives:
\begin{align}
v_1(x_1) &= v(x_1) + (1-x_1) [ -a +(1-b)w ]\;, \label{eq129a} \\ 
v_2(x_1) &= v(x_1) - x_1 [ -a +(1-b)w ]\;, \label{eq129b}
\end{align}
In particular, for $x_1 = 0$:
\begin{align}
v_1(0) &= v_2^* - a +  (1-b) w(0) \;,\label{eq3002g} \\
v_2(0) &= v_2^*\;. \label{eq3002h}
\end{align}
Likewise, setting $x_1=1$ gives
\begin{align}
v_1(1) &= v_1^* \;, \\ 
v_2(1) &= v_1^* + a - (1-b) w(1)\;. \label{eq3002k}
\end{align}
We use equations (\ref{eq3002g}) and (\ref{eq3002k}) to find
\begin{align}
a &= \frac{[v_2^*-v_1(0)]w(1)+[v_1^*-v_2(0)]w(0)}{w(1)-w(0)}\;.\label{eq3003a}\\
b &= 1 - \frac{(v_1^* + v_2^*) - [v_1(0) + v_2(1)]}{w(1)-w(0)}\;,\label{eq3003b}
\end{align}
We see that in order to determine the specific volumes for the two species $v_1(x_1)$ and $v_2(x_1)$, we need to know 
\begin{enumerate}
\item the molar volume $v(x_1)$, and
\item the four endpoints of the two specific volumes $v_1(0)$, $v_1(1)=v_1^*$, $v_2(0)=v_2^*$, and $v_2(1)$. 
\end{enumerate}

\section{Molecular dynamics simulations}
\label{sec:md}

We used a binary mixture of LJs particles in a cubic simulation box with periodic boundary conditions. 
The LJs potential is a smoothly truncated Lennard-Jones potential that was first used by Holian and Evans for simulations of viscosity \cite{holian1983}, see also \cite{hafskjold2019thermodynamic} for thermodynamic properties.
It is equal to the Lennard-Jones (LJ) potential in the range $0 < r < r_s$, where $r_s$ is the LJ potential's inflection point.
In the range $r_s < r < r_c$, it is a third order polynomial and for $r>r_c$,it is zero:
\begin{small}
\begin{equation}
u_{\text{LJ/s}}(r)=
  \begin{cases}
  4\epsilon\left[ \left( \dfrac{\sigma}{r}\right)^{12}-\left( \dfrac{\sigma}{r}\right)^{6}  \right]  & \text{if }r \leq r_s \\ 
  \alpha(r-r_c)^2+\beta(r-r_c)^3 & \text{if }r_{s}<r \leq r_c  \\ 
  0 & \text{if }r>r_c
  \end{cases}
  \label{eqn:ljs}
\end{equation}
\end{small}
The $\epsilon$ and $\sigma$ are the Lennard-Jones energy and size parameters.
The parameters $\alpha$, $\beta$, and $r_c$ are determined such that the potential and its derivative are continuous at $r_s$ and $r_c$.
This means that the force is zero at $r_c$ and so the delta-function contribution to the force in a truncated LJ potential is avoided.
The spline parameters are $r_s=\left ( \frac{26}{7} \right ) ^{(1/6)} \sigma \approx 1.24 \sigma$, $r_c=\frac{67}{48} r_s \approx 1.74 \sigma$, $\alpha =-\frac{24192}{3211} \frac{\epsilon}{r_s^2}$, and $\beta =-\frac{387072}{61009}\frac{\epsilon}{r_s^3}$. 

Three binary mixtures with different potential parameters were used, see Table~\ref{parameters}.
The potential parameters were selected to investigate different deviations
from ideality.

\begin{table}[htb]
\centering
\caption{Potential parameters for the three LJs mixtures.\\
\textnormal{The symbols $\sigma_{ij}$ and $\epsilon_{ij}$ represent parameters in the Lennard-Jones potential.}}
\begin{tabular}{c|cccc}
Mixture & 1 & 2 & 3 \\
\hline
$\sigma_{11}$ & 1.0 & 1.0 & 1.0\\
$\sigma_{12}$ & 1.0 & 1.3 & 1.3\\
$\sigma_{22}$ & 1.0 & 1.0 & 1.3\\
$\epsilon_{11}$ & 1.0 & 1.0 & 1.0\\
$\epsilon_{12}$ & 1.5 & 1.0 & 1.3\\
$\epsilon_{22}$ & 1.0 & 1.0 & 1.3\\
\end{tabular}
\label{parameters} 
\end{table}

All MD simulations were carried out with LAMMPS~\cite{LAMMPS2022}. 
We initialized the LJs mixtures on a regular grid, corresponding to a 
reduced density of $0.5$. To create a system with a specified mole fraction, we first initiated all particles as type 1. A fraction of these particles were randomly selected and converted to type 2 to match the specified mole fraction. In all cases, the
total number of particles was 16,384 and the simulation time step 
was $\tau=0.002$ (LJ units).

To equilibrate the systems at the desired reduced temperature ($T^* = 2$)
and pressure ($p^* = 2$), we first performed NVT simulations lasting 100,000 time steps to achieve the target temperature. Initial velocities were drawn from a Maxwell-Boltzmann distribution (corresponding to $T^* = 2$), and we applied a Nose-Hoover thermostat with a damping constant of $100\tau$ to maintain the temperature. 
After the NVT simulations, we equilibrated the systems further under NPT conditions, where the temperature and pressure were maintained at their target values using a Nose-Hoover thermostat and barostat with damping constants of $100\tau$ and $1000\tau$, respectively, for the temperature and pressure. 
These equilibration NPT simulations were performed for $3\times10^6$ time steps.
After the equilibration phase, we ran production NPT simulations for $5\times10^6$ steps. 
We sampled volume and densities every 10th step and collected system snapshots (for the specific volume calculation) every 5,000 time steps.
The less frequent sampling of snapshots was selected to reduce file sizes and the computational load of the Voronoi tesselation while maintaining accuracy.

For all three mixtures, we have considered 19 compositions from $x_1 = 0.05$ to $x_1 = 0.95$ (with a spacing of $0.05$), and six additional compositions at $x_1 = \{0, 0.01, 0.025, 0.975, 0.99, 1 \}$ for a total
of 25 compositions.

The Redlich-Kister models and the polynomial expansions in Eq.~(\ref{eq130f}) were fitted using least squares regression. The number of terms (the polynomial order) was selected via leave-one-out cross-validation, by minimizing the root-mean squared error of prediction (RMSEP) as a function of the polynomial order. The significance of individual model coefficients was assessed using $t$-tests.
\section{Voronoi tesselation and specific volumes}
\label{sec:voronoi}

To compute specific volumes, we used P-tessellation~\cite{kadtsyn2022volumetric}
as implemented in the Python library \textit{Tess}~\cite{tess}, an interface to \textit{voro++}~\cite{voro2009}. As discussed by Kadtsyn et al.~\cite{kadtsyn2022volumetric}, various mathematical definitions exist for partitioning space around spheres, and P-tessellation is a computationally efficient approximate method that takes into account the different radii of the particles. We defined the particle radii as $\sigma_{11}/2$ and $\sigma_{22}/2$ for the two particle types, respectively.

When applied to a configuration from the MD simulations, the method assigns a volume to each particle in the simulation. We averaged the individual atomic volumes over particle types and 1,000 configurations from each simulation. The standard deviation for the computed specific
volumes were less than $2.2\%$ in all cases.

\section{Results}
\label{sec:results}

\begin{figure}
\centering
\includegraphics[trim=150 70 150 90, clip, width=1.0\linewidth]{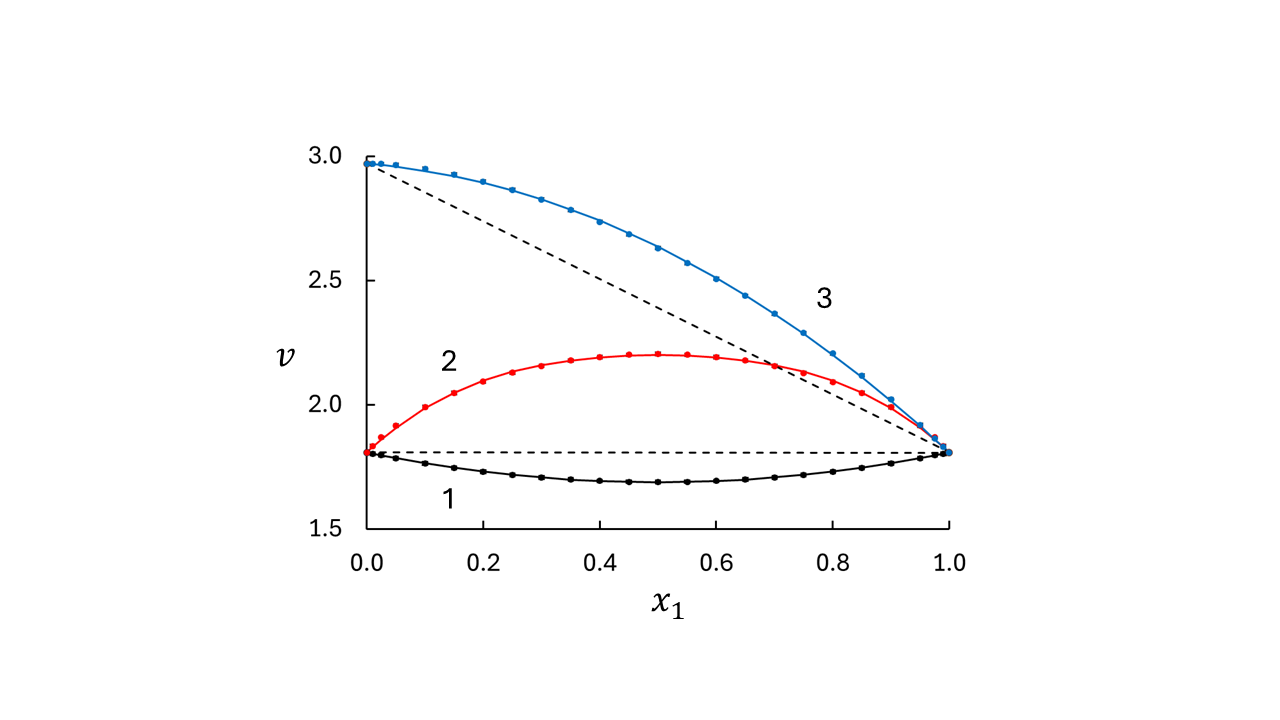}
\caption{Molar volumes  ($v$) as function of  mole fraction of component 1 ($x_1$) for mixtures 1 - 3. The dots are the MD results, the dashed lines are the ideal volumes of mixing, and the solid lines are fitted Redlich-Kister model. The error bars in the MD data are represented by the size of the symbols.}
\label{fig:molar_volumes}
\end{figure}

Molar volumes for the three mixtures are shown as function of mole fraction $x_1$ in Figure~\ref{fig:molar_volumes}.
Mixture 1 has a negative deviation from ideal mixing volume due to the stronger attraction between unlike particles ($\epsilon_{12}>\epsilon_{11}=\epsilon_{22}$). 
Mixture 2 has a positive deviation from ideal mixing volume due to the larger apparent diameters between unlike particles ($\sigma_{12}>\sigma_{11}=\sigma_{22}$).
The pure components in mixtures 1 and 2 are identical, hence their molar volumes are equal.  Mixture 3 has a combination of particle-size and -energy effects due to the different $\epsilon$- and $\sigma$-values between the pure components and between the unlike particles. Mixture 3 shows a positive deviation from ideal mixture.


The first few parameters in the Redlich-Kister expansion were fitted to the simulation data, giving the results listed in Table \ref{RK_coefficients}. There was no significant reduction of RMSEP when increasing the polynomial order beyond the first term for mixture 1. For mixtures 2 and 3, a significant reduction in RMSEP was obtained for third-order models (with no significant reduction for higher orders). Based on this, we used one term in the Redlich-Kister model for mixture 1 and three terms for mixtures 2 and 3. In all cases, the adjusted coefficient of determination, $R^2_\text{adj}$, was greater than $0.999$.
The lines in Figure~\ref{fig:molar_volumes} show the fitted data.

\begin{table}[htb]
\centering
\caption{Redlich-Kister parameters for the three LJs mixtures. \textnormal{Errors are estimated as twice the standard error of the parameters. Parameters not significantly different 
from zero (at a significance level of 0.01) are marked with an asterisk ($\ast$). For mixture 1, only one term was needed in Redlich-Kister expansion (see Section \ref{sec:results}); solid lines indicate the parameters that were not obtained.}}
\begin{tabular}{c|ccc}
Mixture & 1 & 2 & 3 \\
\hline
$A_0$ & $-0.4783 \pm 0.0005$ & $1.567 \pm 0.015$ & $0.9595 \pm 0.0016$\\
$A_1$ & --- & $-0.0004 \pm 0.03$$^\ast$ & $0.025 \pm 0.003$ \\
$A_2$ & --- & $0.66 \pm 0.07$& $0.171 \pm 0.007$ \\
\end{tabular}
\label{RK_coefficients} 
\end{table}

The partial molar volumes were computed from analytic differentiation of the Redlich-Kister expansion.

The specific (Voronoi) volumes were computed from the same simulations as the molar volumes.
Selected configurations were analysed with Voronoi tesselation as described in Section \ref{sec:voronoi}.

\begin{figure}
\centering
\includegraphics[trim=150 70 150 70, clip,width=1.0\linewidth]{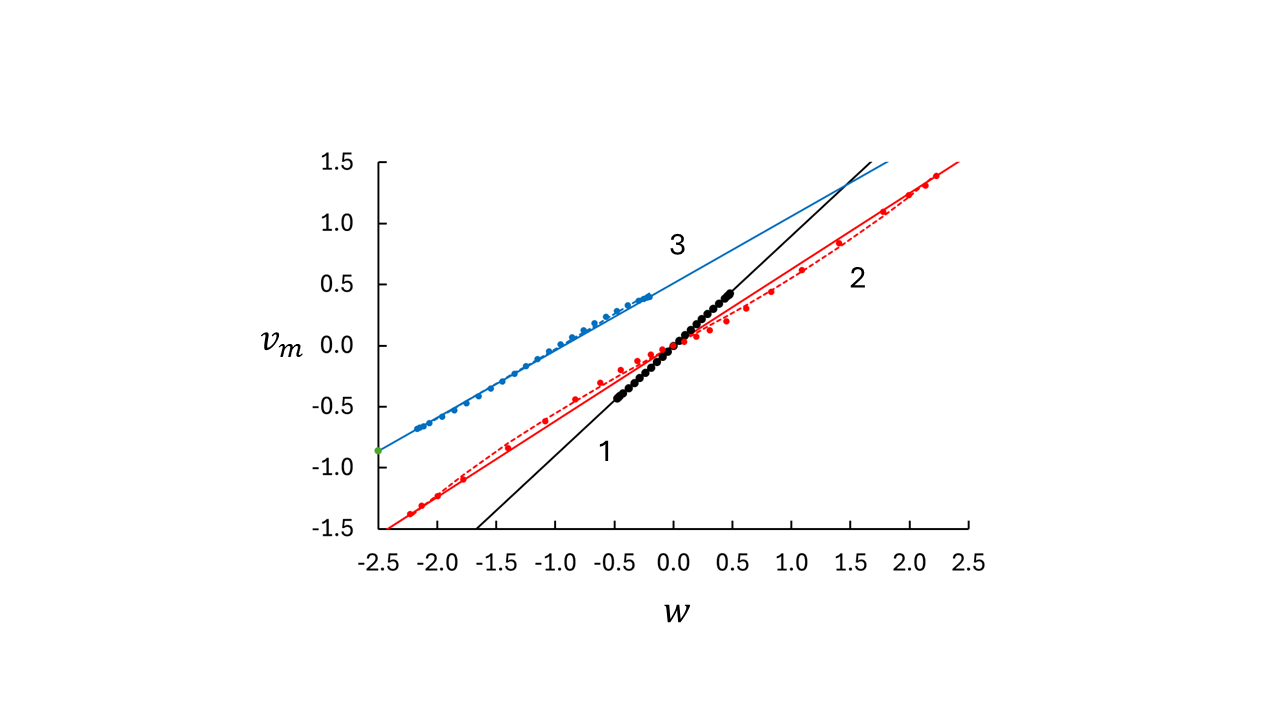}
\caption{Co-molar volume ($v_m$) as a function of partial molar volume difference ($w = \hat v_1 - \hat v_2$) for the three mixtures. 
The dots were obtained from simulations and the straight solid lines from Eqs. (\ref{eq3003a}) and (\ref{eq3003b}). 
The dashed line for mixture 2 represents a third order polynomial fitted to the simulation data (see text).
}
\label{fig:vm_vs_w}
\end{figure}

Figure \ref{fig:vm_vs_w} shows $v_m$ as function of $w$ for the three mixtures.
The extrapolated values $v_1(0)$ and $v_2(1)$ were determined from polynomial fits to the MD results.

We found that $v_m$ is a perfect linear function of $w$ for mixtures 1 and 3, and a very good approximation for mixture 2.
For mixture 2, the function is linear in the central part, but with a slight S-shaped curve overall.
Using Eqs. (\ref{eq3003a}) and (\ref{eq3003b}) gave the parameter values listed in the upper part of Table~\ref{vm_coefficients}.
When we fitted polynomials to the MD data, using Eq.~(\ref{eq130f}), we found the parameters listed in the lower part of Table~\ref{vm_coefficients}. First order polynomials gave precise fits for all mixtures ($R^2_{\text{adj}} = 0.9999$, $0.997$, and $0.9991$, respectively). Still, for mixture 2, we used a third-order polynomial to demonstrate the possible non-linearity of the relationship in Eq.~(\ref{eq130f}).

\begin{table}[htb]
\centering
\caption{Parameters for $v_m$ as function of $w$. \textnormal{The upper part shows $a$ and $b$ determined from Eqs.~(\ref{eq3003a}) and~(\ref{eq3003b}). The lower part shows fitted values of $a_k$ from Eq.~(\ref{eq130f}) as described in the Materials and Methods.
Errors are estimated as twice the standard error of the parameters. Parameters not significantly different from zero (at a significance level of 0.01) are marked with an asterisk ($\ast$).
For Mixtures 1 and 3, only first-order polynomials were fitted; lines indicate that the corresponding parameters were not obtained.}}
\begin{tabular}{c|cccc}
Mixture & 1 & 2 & 3 \\
\hline
$a$ &$-0.0011$ & $-0.0045$ & $0.5113$ \\
$b$ & $0.8990$ & $0.6217$ & $0.5492$ \\
\hline
$a_0$ & $0.0001 \pm 0.001^\ast$  & $-0.0001 \pm 0.02^\ast$  &  $0.570 \pm 0.010$ \\
$a_1$ & $0.906 \pm 0.004$ & $0.53 \pm 0.03$ &  $0.595 \pm 0.007$ \\
$a_2$ & ---  & $0.0004 \pm 0.006^\ast$ &  --- \\
$a_3$ & ---  & $0.019 \pm 0.006$ & --- \\
\end{tabular}
\label{vm_coefficients} 
\end{table}

\section{Discussion and conclusions}
\label{sec:fin}

\begin{figure}
\includegraphics[trim=150 70 150 70, clip,width=1.0\linewidth]{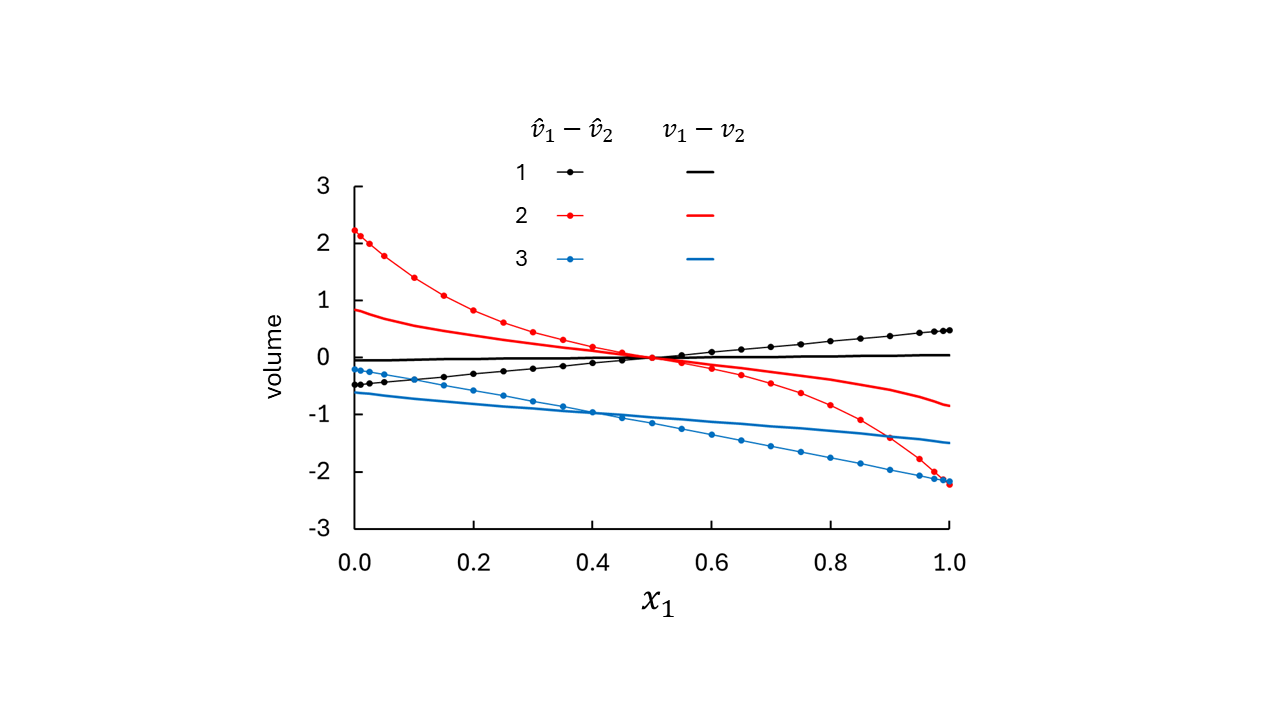}
\caption{Differences between partial molar volumes ($\hat{v}_1 - \hat{v}_2 = w$) and specific volumes ($v_1 - v_2$) as a function of $x_1$ for the three mixtures. 
Note that the behavior is linear for mixtures 1 and 3, and curved for mixture 2.
}
\label{fig:volume_differences}
\end{figure}

In order to get a better understanding of the relationship between $v_m$ and $w$, we examined the differences $\hat{v}_1 - \hat{v}_2 (= w)$ and $v_1 - v_2$ as a function of $x_1$. 
The results are shown in Figure \ref{fig:volume_differences}.

\begin{figure}
\centering
\includegraphics[trim=150 70 150 70, clip,width=1.0\linewidth]{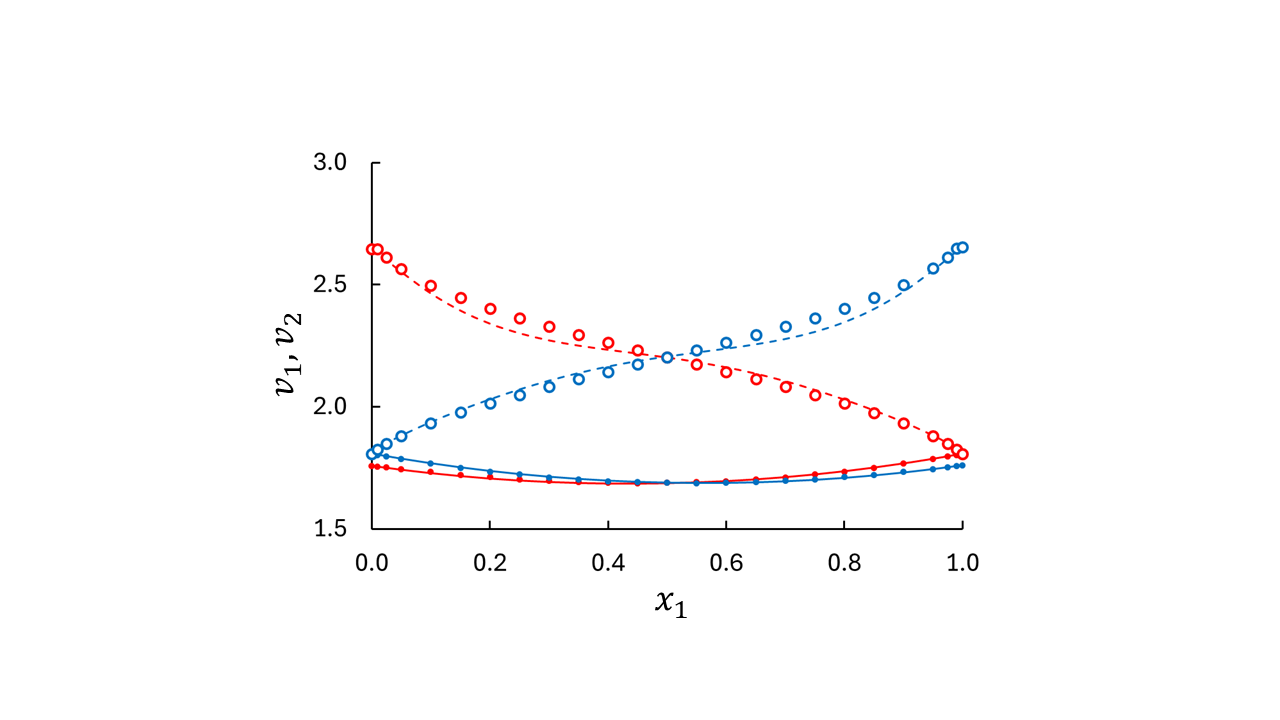}
\caption{Back calculation of $v_1$ and $v_2$. The solid circles and lines (bottom) show results for mixture 1, the open circles and dashed lines (top) show results for mixture 2. The red and blue colors represent $v_1$ and $v_2$, respectively. The circles and lines represent results from the Voronoi tesselation and back-calculated result, respectively.
}
\label{fig:back}
\end{figure}

Mixture 2 behaves differently from the other two in that both $v_1 - v_2$ and $w$ are non-linear, especially for compositions near the pure components.
The linearity of both functions for the other two mixtures give the linear plots shown in Figure \ref{fig:vm_vs_w}.
To further examine how the non-linear behavior for mixture 2 affects the major topic of this work, namely the possibility of extracting partial molar volumes from the specific volumes and vice versa, we use the linear approximation for $v_m(w)$ to back-calculate $v_1$ and $v_2$ from Eqs. (\ref{eq129a}) and (\ref{eq129b}).
The results are shown in Figure \ref{fig:back}.
Results for mixture 3 are not shown because they are similar to the mixture 1 results.

The overall result is that a linear relation between $v_m$ and $w$ does an excellent job for mixtures 1 and 3 and a reasonable job for mixture 2, which was the most difficult case of the mixtures considered here.

The analysis of the co-molar volume depends critically on estimates of the parameters $a$ and $b$ in Eqs. \eqref{eq3003a} and \eqref{eq3003b}.
In this work, we have used the extrapolated values of the functions $v_1(x_1)$ and $v_2(x_1)$ (the other variables in these equations can be obtained from molar volumes).
Our final objective is to determine the specific volumes from the partial molar volumes without relying on extrapolations of $v_1(x_1)$ and $v_2(x_1)$.
We are currently exploring alternative methods to achieve this.

To conclude, we have established a theory that relates the classical and measurable partial molar volumes for binary miscible mixtures to the more intuitive specific (Voronoi) volumes that can be computed by simulation.
The theory was verified by MD simulations of both kinds of volumes for three mixtures of Lennard-Jones/spline particles.
Specifically, we have compared the two partial molar volumes $\hat{v}_1$ and $\hat{v}_2$ (equations (\ref{eq4}) and (\ref{eq5})) to the two specific volumes $v_1$ and $v_2$ (equations (\ref{eq7}) and (\ref{eq8})).

This theory has given rise to three surprises:
\begin{itemize}
\item The two pairs of volumes are related through a {\it single\/} function $v_m$ --- the co-molar volume,
\item the co-molar volume can be expressed using the specific volumes $v_1$ and $v_2$ {\it alone,\/}
see equation (\ref{eq3002b}), and lastly
\item the co-molar volume is to a good approximation {\it linear\/} in the natural variable $w$, defined in equation
(\ref{eq7a}), see equation (\ref{eq130c}).
\end{itemize}
We have with the introduction of the co-molar volume created a two-way mapping between the molar volume and the co-molar volume on one hand, and the specific molar volumes on the other hand, \textit{i.e.}, $(v,v_m) \rightleftharpoons (v_1,v_2)$. 
Equations (\ref{eq129a}), (\ref{eq129b}), and (\ref{eq130f}) provide the mapping $(v,v_m) \rightarrow (v_1,v_2)$:
\begin{align}
v_1 =& v + x_2 \left ( w- \sum_{k=0}^\infty a_k w^k \right ) \\
v_2 =& v - x_1 \left ( w- \sum_{k=0}^\infty a_k w^k \right ) \;.
\end{align}
The inverse mapping $(v,v_m) \leftarrow (v_1,v_2)$ is given by equations (\ref{eq19a}) and (\ref{eq3002b}):
\begin{align}
v_m =& x_1 \left(\frac{\partial v_1}{\partial x_1} \right )_{T,p} + x_2 \left(\frac{\partial v_2}{\partial x_1} \right )_{T,p} \\
w =& v_m + (v_1 - v_2) \;.
\end{align}

We have here focused on the volume of mixtures.  However, this is but one quantity.  We expect that any quantity that characterizes properties of mixtures may be accompanied by a corresponding co-property function as we have here defined the molar volume and the co-molar volume.  
This makes it possible to  incorporate the directly measured corresponding partial quantity for each species into the thermodynamic formalism of mixtures. 
An example of such a quantity is the potential energy of the mixture.  

Our introduction of the co-molar volume was inspired by recent work done in a very different context, namely immiscible two-phase flow in porous media. Hansen et al.\ \cite{hansen2018relations} introduced a formalism that in mathematical structure is identical to the thermodynamics of mixtures, but dealing with flow velocities.  In order to derive from this formalism the velocity of each fluid species, they introduced the concept of a {\it co-moving velocity\/} \cite{roy2020flow,roy2022co,pedersen2023parameterizations,hansen2023statistical,alzubaidi2024impact,hansen2024linearity,feder2022physics}.  The co-moving velocity is analogous to the co-molar volume introduced here, and it behaves in the same linear way with respect to the variable corresponding to $w$. Finding an analog in {\it ordinary thermodynamics\/} to the co-moving velocity with the same properties, lends strong support to the thermodynamics-like approach to flow in porous media.   
\bigskip\bigskip


This work was partly supported by the Research Council of Norway through its Center of Excellence funding scheme, project number 262644. Further support, also from the Research Council of Norway, was provided through its INTPART program, Project No. 309139. AH acknowledges funding from the European Research Council (Grant Agreement 101141323 AGIPORE).
\bigskip\bigskip

\bibliography{references}
\end{document}